\documentclass[12pt]{article}
\usepackage[intlimits,tbtags]{amsmath}
\usepackage{amsfonts,amssymb,amsthm}
\usepackage[latin1]{inputenc}
\usepackage[usenames,dvips]{color}

\def\beq{\begin{equation}}
\def\eeq{\end{equation}}
\def\bea{\begin{eqnarray}}
\def\eea{\end{eqnarray}}
\def\nn{\nonumber}

\begin{document}

\topmargin -0.5cm \oddsidemargin -0.8cm \evensidemargin -0.8cm
\pagestyle{empty}

\begin{center}
\vspace*{5mm}
{\bf A realization of the quantum Lorentz group} \\
\vspace*{0.4cm}
{\bf Boyka Aneva} \\
\vspace*{0.2cm}
{Theory Division, CERN, 1211 Geneva 23, Switzerland}\\
\vspace*{0.6cm}
{\bf Abstract} \\
\end{center}
A realization of a deformed Lorentz algebra is considered and its
irreducible representations are found; in the limit $q\rightarrow
1$, these are precisely the irreducible representations of the
classical Lorentz group.

\setcounter{page}{1} \pagestyle{plain}

Since the invention of the quantum group as a pure mathematical
structure \cite {drin} much progress has been made in the
realization of deformations of simple Lie algebras \cite {jim},
and also Lie superalgebras, and in the construction of their
representations. Quantization of space-time symmetry groups
(Lorentz, Poincare, the conformal group) has remained a problem
for incorporating ana applying the structure of quantum groups
into physical systems. A deformation of the Lorentz group has been
studied in \cite {podl} and a six-generator deformed Lorentz
algebra has been found in \cite {ogw} in terms of the chiral
$SL(2)\times SL(2)$ generators.

In this paper we consider a realization of deformed Lorentz
algebra commutation relations obeyed by the generators - rotations
and boosts - and construct its representations which are the exact
quantum analogue of the classical Lorentz group representations.

The quantum group is generally defined as a $q$-deformation of the
universal enveloping algebra of the underlying classical group.
Thus when introducing a quantum group one should first construct a
deformed associative with a non-cocommutative Hopf algebra
structure; this is usually done in terms of generators obeying
deformed Lie commutation relations.

We recall that the Lorentz group contains the  $SU(2)$ subgroup of
rotations $M_i, i=1,2,3$, and the boost generators $N_i, i=1,2,3$
form an irreducible $SU(2)$  vector representation. One can form
two chiral $SL(2)$ subgroups $I_i^L=M_i+iN_i$ and $I_i^R=M_i-iN_i$
which act only on spinors with undotted and dotted indices. The
rotation $SU(2)$ subgroup is the diagonal in $SL^L(2)\times
SL^R(2)$. The matrix elements of the two-dimensional fundamental
representation generators (and of the conjugated one) satisfy
$\langle N_i\rangle=\mp i\langle M_i\rangle$, the latter
expressing that $I^{L,R}_i$ vanish when acting on functions of
only dotted and undotted indices respectively.

In defining a deformed quantum Lorentz group we wish to generalize
the properties of the classical Lorentz group to the $q$-case. We
shall determine the deformed commutation relations as relations
imposed on the matrix elements of the fundamental representation
generators acting on two-dimensional spinors with undotted
indices. We assume a full analogy with the classical case, namely
that the quantum Lorentz group contains the $SU_q(2)$ subgroup,
the boost operators transform under $SU_q(2)$ as components of
irreducible tensor operators, and the fundamental representation
has the properties of the two-dimensional classical spinor
representation.

Let $M_{\pm}, M_3$ satisfy the deformed commutation relations of
$SU_q(2)$ \cite {bied1}. In the deformed case the $SU_q(2)$
generators do not transform under the irreducible tensor
representation. The $q$-analogue of an irreducible $SU_q(2)$
tensor operator has been defined \cite {scheu, smir} as the set of
$2l+1$ components $T^l_m$, $m=-l,...,l$ obeying \bea
[M_3,T^l_m]&=&mT^l_m,
\\  \nn [M_{\pm}, T^l_m]_{q^{-m/2}}&=&[l\mp m]^{1/2}[l\pm m
+1]^{1/2}T^l_{m\pm 1}q^{M_3/2} \label{} \eea where
$[A,B]_q^{\alpha}=AB-q^{\alpha} BA$. There is an alternative
definition to (1), obtained by $q\rightarrow q^{-1}$. Accordingly
one can construct two $q$-vector operators $S$ and $T$: \bea
S_{\pm} = \pm q^{\mp}M_{\pm}q^{-M_3/2}  \\  \nn
S_0=[2]^{-1/2}(q^{-1/2}M_-M_+-q^{1/2}M_+M_-) \label{} \eea and
\bea T_{\pm}=\pm q^{\pm}M_{\pm}q^{M_3/2}  \\  \nn
T_0=[2]^{-1/2}(q^{1/2}M_-M_+-q^{-1/2}M_+M_-) \label{} \eea We
identify the operators $-iq^{-1/4}M_+q^{-M_3/2},
-iq^{-1/4}M_-q^{M_3/2}$ and $-i[M_3]q^{-M_3/2}$ with the
generators $N_+, N_-$ and $N_3$respectively,  of a two-dimensional
$q$-deformed Lorentz boost transformation, expressed in
a$q$-tensor form. We identify further the operators $M_{\pm}$ in a
$q$-tensor form, i.e. $M_{\pm}\equiv q^{-1/4}M_{\pm}q^{\mp}M_3/2$,
and the operator $M_3$ with the rotation generators of a
$q$-deformed Lorentz transformation. The commutation relations of
a deformed Lorentz algebra are imposed as the relations obeyed by
the generators of the two-dimensional Lorentz transformation.
Namely, the rotations $M_{\pm}, M_3$ satisfy the Lie brackets of
the deformed $U_q(su(2))$; the commutation relations between
rotations and boosts are determined as the action of the
$U_q(su(2))$ generators on the irreducible $su_q(2)$ tensor
operators (2) and (3); the commutators between the boosts are
such, that in the limit $1\rightarrow 1$ the classical
two-dimensional boost generators are recovered. The deformed
Lorentz algebra has the form \bea M_+M_--M_-M_+&=&[2M_3], \\  \nn
M_3M_{\pm}-M_{\pm}M_3&=&\pm M_\pm, \\ \nn N_+N_--N_-N_+&=&-[2M_3] \\
\nn N_3N_+q^{1/2}-q^{-1/2}N_+N_3&=&-M_+, \\  \nn \tilde
{N_3}N_-q^{1/2}-q^{-1/2}N_-\tilde {N_3}&=&M_-,  \\  \nn
M_3N_{\pm}-N_{\pm}M_3&=&\pm N_{\pm}, \\  \nn
M_+N_-q^{-1/2}-q^{1/2}N_-M_+&=&[2]\tilde {N_3}
+(q^{1/2}-q^{-1/2})C'^q_2, \\  \nn
M_-N_+q^{-1/2}-q^{1/2}N_+M_-&=&-[2]N_3
+(q^{1/2}-q^{-1/2})C'^{q}_2,
\\  \nn M_+\tilde {N_3}q^{1/2}-q^{-1/2}\tilde {N_3}M_+&=&-N_+, \\  \nn
M_-N_3q^{1/2}-q^{-1/2}N_3M_-&=&N_- \label{} \eea and all other
(usual) commutators vanish. The element \beq 2C'^{q}_2=\frac
{M_+N_-q^{-1/2}-N_-M_+q^{1/2}+M_-N_+q^{-1/2}-N_+M_-q^{1/2}}{q^{1/2}-q^{-1/2}}
\label{}\eeq is central in the algebra and is the quantum analogue
of the second order Lorentz Casimir $C'_2=-M_iN_i, i=1,2,3$. In
the limit $q\rightarrow 1$ the deformed Lorentz algebra (4)
contracts to the Lie algebra of the classical Lorentz group. One
can show that the operators $M_{\pm}, M_3, N_{\pm}(q^{-1}), N_3$
satisfy the $q$-deformed Lorentz algebra (4) with $q$ replaced by
$q^{-1}$ and $N\leftrightarrow \tilde{N_3}$. Then a $q$-adjoint
involution on the deformed Lorentz group can be
defined \bea (M_{\pm})^*&=&M_{\mp}, \quad\quad\quad M_3^*=M_3, \\
\nn (N_{\pm}^*&=&N_{\mp}(q^{-1}), \quad N^*_3=N_3 \label{} \eea

The $q$-deformed Lie-brackets (4) and the corresponding ones with
$q$ replaced by $q^{-1}$ and $N_3\leftrightarrow \tilde N_3$
define a six-generator quantum Lorentz group.

We note an interesting novelty, the appearance of the central
element in the defining commutation relations of the deformed
Lorentz algebra (16). It is a property of the deformed relations
obeyed by the generators of the  quantized universal enveloping
algebra which  is an associative algebra with a unit and with a
Poincare-Birkoff-Witt basis.

To construct the irreducible representations of the deformed
Lorentz algebra we follow the classical procedure. Namely, the
representations are realized in the space of the $U_q(su(2))$
irreducible representations with canonical basis $\vert j,
m\rangle_q$ with $j$ integer or half integer and $m=-j,...,j$. The
action of the rotation and boost generators on the basis vectors
is given by: \beq M_{\pm} \vert j, m\rangle_q = [j\mp
m]^{1/2}[j\pm m+1]^{1/2}q^{-1/4}q^{\mp m/2}\vert j, m\pm
1\rangle_q \label{} \eeq \beq M_3\vert j, m\rangle_q=m\vert j,
m\rangle_q \label{}\eeq \bea N_{\pm}\vert j, m\rangle_q&=&\pm
c_j[j\mp m]^{1/2}[j\mp m-1]^{1/2}q^{-1/4}q^{-(j\pm m)/2}\vert j-1,
m\pm \rangle_q \\ \nn &-&a_j[j\mp m]^{1/2}[j\pm
m+1]^{1/2}q^{-1/4}q^{\mp}m/2\vert j, m\pm 1\rangle_q \\ \nn &\pm&
c_{j+1}[j\pm m+1]^{1/2}[j\pm m+2]^{1/2}q^{1/4}q^{(j\mp m)/2}\vert
j+1, m\pm 1\rangle_q\label{}\eea \bea N_3\vert j, m\rangle_q
&=&c_j[j-m]^{1/2}[j+ m]^{1/2}q^{-m/2}\vert j-1, m\rangle_q \\ \nn
&-&a_j[m]q^{-m/2}\vert j, m\rangle_q  \\  \nn
&-&c_{j+1}[j+m+1]^{1/2}[j-m+1]^{1/2}q^{-m/2}\vert j+1, m\rangle_q
\label{}\eea The action of the operator $C'^q_2$ on the basis
vectors is given by \beq C'^q_2\vert j, m\rangle_q
=i[l_0][l_1]\vert j, m\rangle_q \label{}\eeq The coefficients
$a_j, c_j$ can be determined by using the commutators between the
generators $N_+$ and $N_-$ and $N_\pm$, and $N_3$ which results in
the pair of difference equations \beq
(a_{j+1}[j+2]-a_j[j])c_{j+1}=0, \label{}\eeq \beq
c_j^2[2j-1]-a_j^2-c_{j+1}^2[2j+3]=1 \label{}\eeq We first note
that since $j\geq 0$ there is a minimal (integer or half integer)
value $j_{min}=l_0$ and hence $j=l_0+1,l_0+2,...$. Assuming that
the coefficient $c_{l_0}=0$ we have two possibilities, either \beq
c_{l_0}=0, \quad c_{l_0+1}\neq 0, ...c_{l_0+n}\neq 0,\quad
c_{l_0+n+1}=0,\label{}\eeq and the representation is
finite-dimensional, or \beq c_{l_0}=0, \quad c_j \neq 0 \quad for
\quad any \quad j>l_0, \label{}\eeq and the representation is
infinite-dimensional. Eqs.(12, 13) for the coefficient can be
easily solved and the result, being dependent on two constants
$l_0, l_1$ is \beq a_j=\frac {i[l_0][l_1]}{[j][j+1]}, \quad
c_j=\frac {i}{[j]}\sqrt {\frac
{([j]^2-[l_0]^2)([j]^2-[l_1]^2)}{[2j-1][2j+1]}}\label{}\eeq for
any $j>l_0$. Since $j=l_0+n$, where $n$ is a natural number, the
representation will be finite-dimensional if for some $n$ \beq
[l_1]^2=[l_0+n+1]^2\label{}\eeq Due to the property of the
quantity $[A]$ the above equation is satisfied for $\pm l_1=\pm
(l_0+n+1)$. The parameter $l_1$ is in general a complex number,
but $l_0+n+1$ is a real positive number, so that the
representation series will terminate if, for some real $l_1$, \beq
\vert l_1\vert =l_0+n+1. \label{}\eeq hence the spin content of
the irreducible finite-dimensional representation Lorentz
$q$-representation is determined by \beq j=l_0, l_0+1,...,\vert
l_1\vert -1, \qquad m=-j, -j+1,...,j. \label{}\eeq

We summarize the result: The irreducible representation of the
deformed Lorentz algebra is determined by the pair $([l_0],[l_1]$,
where $l_0$ is a non-negative integer or half-integer real number
and $l_1$ is a complex number. The irreducible representation
corresponding to a given pair $([l_0], [l_1])$ in the $U_q(su(2))$
canonical basis $\vert j, m\rangle_q$ is given by formulae (7-11)
with the coefficients (16).\\ If, for some natural number $n$,
\beq [l_1]^2=[l_0+n+1]^2 \label{}\eeq then the representation is
finite-dimensional with the possible values of $j$ and $m$ given
by (19). If \beq [l_1]^2\neq [l_0+n+1]^2, \label{} \eeq then the
representation is infinite-dimensional. In the limit $q\rightarrow
1$, (7-11) and (16) reproduce exactly the irreducible Lorentz
group \cite {nai} representations.

We now consider the conditions under which the representations of
the deformed Lorentz algebra are unitary. Since the generator
$N_3$ is self-q-adjoint, according to (6), then \bea \langle
j,m\vert N_3 \vert j,m\rangle &=& \langle j,m\vert N_3^* \vert j,m \rangle, \\
\nn \langle j-1,m\vert N_3 \vert j-1,m \rangle &=& \langle j-1,m
\vert N_3^* \vert j-1,m\rangle \label{} \eea Hence $a_j=\bar{a_j}$
and $c_j=-\bar{c_j}$. From the first of the formulae (16), it
follows that the condition $a_j=\bar{a_j}$ is satisfied if either
$[l_1]$ is arbitrary and $[l_0]=0$, or $[l_0]=0$ is arbitrary and
$i[l_1]$ is real. The second possibility with $q$ real means that
$l_1$ should be pure imaginary \beq l_1=i\rho, \label{}\eeq with
$\rho$ real. The condition $c_j=-\bar{c_j}$ for the second of the
formulae (16) means that the expression under the square root must
be positive, and this is obviously the case if, only \beq
[j]^2-[l_1]^2>0 \label{}\eeq We have to consider two possibilities:\\
(a) $l_0\neq 0$ and $[l_1]$ pure imaginary, which coincides with
(23).\\
(b) $[j]^2\geq [l_1]^2$ with $[l_1]$ real.\\
The latter expression has to be satisfied for all $j$ and this is
only possible if $[l_1]^2\leq [1]$. Hence the possible values of
$[l_1]$ are \beq 0<\vert l_1 \vert \leq 1 \label{}\eeq The
relations between $N_{\pm}$ and their $q$-adjoint yield  the same
values for $l_0$ and $l_1$.

We thus conclude: The irreducible representations of the deformed
Lorentz algebra determined by the pair $[l_0],[l_1]$ is unitary if
either $l_1$ is pure imaginary and $l_0$ is an arbitrary
non-negative integer or half-integer, or $l_0=0$ and $l_1$ is a
real number in the interval $0<\vert l_1 \vert \leq 1$. In the
limit $q\rightarrow 1$ the corresponding representations (7-11)
reproduce exactly the infinite-dimensional Lorentz group
representations \cite {nai} of the principal and complementary
series respectively.

To analyze the Hopf structure of the quantum lorentz group we need
to generalize to the $q$-case the classical picture of forming two
$SL(2)$ groups from Lorentz rotations and boosts. For this purpose
we consider the operators \bea I_{\pm}^L&=&M_{\pm}+iN_{\pm}, \\
\nn
I_{\pm}^R&&M_{\pm}-iN_{\pm}, \\  \nn I_3^L&=&[M_3]q^{-M_3/2}+iN_3, \\
\nn I_3^R&=&[M_3]q^{-M_3/2}-iN_3, \\  \nn
\tilde{I_3^L}&=&[M_3]q^{M_3/2}+i\tilde{N_3} \\  \nn
\tilde{I_3^R}&=&[M_3]q^{M_3/2}-i\tilde{N_3} \label{}\eea These
operators satisfy the algebra \bea
I_+^LI_-^L-I_-^LI_+^L&=&2(I_3^L+\tilde{I_3^L}), \\  \nn
I_3^LI_+^Lq^{1/2}-q^{-1/2}I_+^LI_3^L&=&2I_+^L, \\ \nn
\tilde{I_3^L}I_-^Lq^{1/2}-q^{-1/2}I_-^L\tilde{I_3^L}&=&-2I_-^L, \\
\nn
\tilde{I_3^L}I_+^Lq^{-1/2}-q^{1/2}I_+^L\tilde{I_3^L}&=&2I_+^L, \\
\nn I_3^LI_-^Lq^{-1/2}-q^{1/2}I_-^LI_3^L&=&-2I_-^L,
\\ \nn
I_+^RI_-^R-I_-^RI_+^R&=&2(I_3^R+\tilde{I_3^R}), \\  \nn
I_3^RI_+^Rq^{1/2}-q^{-1/2}I_+^RI_3^R&=&2I_+^R, \\ \nn
\tilde{I_3^R}I_-^Rq^{1/2}-q^{-1/2}I_-^R\tilde{I_3^R}&=&-2I_-^R, \\
\nn
\tilde{I_3^R}I_+^Rq^{-1/2}-q^{1/2}I_+^R\tilde{I_3^R}&=&2I_+^R, \\
\nn I_3^RI_-^Rq^{-1/2}-q^{1/2}I_-^RI_3^R&=&-2I_-^R,
 \label{}\eea The generators $I^L_{\pm},I^L_3,\tilde{I^L_3}$
simply commute with $I^R_{\pm},I^R_3,\tilde{I^R_3}$. It seems that
the algebra (27) has more than six generators. Due to the
relations \bea 1+\alpha \tilde{I_3^L}+\alpha
\tilde{I_3^L}&=&(1-\alpha I_3^L-\alpha I_3^R)^{-1}, \\  \nn
\tilde{I_3^L}-\tilde{I_3^R}&=&(1+\alpha \tilde{I_3^L}+\alpha
\tilde{I_3^L})(I_3^L-I_3^R) \label{}\eea with
$\alpha=(q^{1/2}-q^{-1/2})/2$, the quantum algebra (27) is, in
fact, generated by two raising, two lowering and two diagonal
operators.

There exists a $q$-adjoint involution in the algebra (27), defined
as \bea (I^L_{\pm}(q))^*=I^R_{\mp}(q^{-1}), \\  \nn
I_3^{L*}=I_3^R, \quad\quad \tilde{I_3^{L*}}=\tilde{I_3^R}
\label{}\eea

Denoting the two-dimensional $q$-deformed Lorentz representations
\bea  \vert 1/2, m;l_0=1/2, l_1=3/2\rangle_q&=&\tau_{1/2}, \\  \nn
\vert 1/2, m;l_0=1/2, l_1=-3/2\rangle_q&=&\tilde{\tau_{1/2}},
\label{}\eea we observe that $I^L_{\pm}, I^L_3)$ (respectively
$I^R_{\pm}, I^R_3)$) vanish when acting on $\tilde{\tau_{1/2}}$
(respectively $\tau_{1/2}$).

The algebra (27) is the $q$-analogue of the chiral decomposition
of the classical Lorentz group, which is exactly reproduced in the
limit $q\rightarrow 1$.

We further introduce the shifted diagonal generators \bea
T^{L,R}_3=2-(q^{1/2}-q^{-1/2})I^{L,R}_3, \\  \nn
\tilde{T^{L,R}_3}=2+(q^{1/2}-q^{-1/2})\tilde{I^{L,R}_3}
\label{}\eea which does not change the structure of the algebra
(27). The co-product for the deformed algebra (27) is given by
which amounts to a non-cocommutative Hopf algebra structure.
\bea \Delta (I_+^L)&=&I_+^L\otimes 1+T_3^L\otimes I_+^L, \\
\nn
\Delta (I_-^L)&=&I_-^L\otimes \tilde{T_3^L}+1\otimes I_-^L, \\
\nn
\Delta (I_+^R)&=&I_+^R\otimes \tilde{T_3^R}+1\otimes I_+^R, \\
\nn
\Delta (I_-^R)&=&I_-^R\otimes 1+T_3^L\otimes I_-^R, \\
\nn \Delta (T^{L,R}_3)&=&T^{L,R}_3\otimes T^{L,R}_3, \\  \nn
\Delta (\tilde{T^{L,R}_3})&=&\tilde{T^{L,R}_3}\otimes
\tilde{T^{L,R}_3}. \label{}\eea

To summarize, we have defined a quantized Lorentz algebra and
found that every irreducible classical Lorentz group
representation labelled by $l_0$ and $l_1$ can be $q$-deformed to
an irreducible $q$-representation of the deformed Lorentz algebra
labelled by $[l_0]$ and $[l_1]$. The possible values of $l_0$ and
$l_1$ are exactly the same as in the classical case.

The author is grateful for the support and hospitality of the
Theory Division at CERN where most of this work was completed.
Partial support of the National Foundation for Scientific Research
under contract $\Phi$-11 is acknowledged.

\end{document}